\documentclass{llncs}
\usepackage{version}
\usepackage{bjmp}

\title{Long Multiplication by Instruction Sequences with Backward 
       Jump Instructions}
\author{J.A. Bergstra \and C.A. Middelburg}
\institute{Informatics Institute, Faculty of Science, University of
           Amsterdam, \\
           Science Park~904, 1098~XH Amsterdam, the Netherlands \\
           \email{J.A.Bergstra@uva.nl,C.A.Middelburg@uva.nl}}

\begin{document}
\maketitle

\begin{abstract}
For each function on bit strings, its restriction to bit strings of any 
given length can be computed by a finite instruction sequence that 
contains only instructions to set and get the content of Boolean 
registers, forward jump instructions, and a termination instruction.
Backward jump instructions are not necessary for this, but instruction 
sequences can be significantly shorter with them.
We take the function on bit strings that models the multiplication of 
natural numbers on their representation in the binary number system to 
demonstrate this by means of a concrete example.
The example is reason to discuss points concerning the halting problem 
and the concept of an algorithm.
\begin{keywords} 
single-pass instruction sequence, backward jump instruction, 
bit string function, long multiplication, algorithm, halting problem,
indirect addressing.
\end{keywords}%
\begin{classcode}
F.1.1, F.2.1.
\end{classcode}
\end{abstract}

\section{Introduction}
\label{sect-intro}

In~\cite{BM13a}, an approach to non-uniform complexity is presented 
which is based on the simple idea that, for each function on bit 
strings, its restriction to bit strings of any given length can be 
computed by an instruction sequence that contains only instructions to 
set and get the content of Boolean registers, forward jump instructions, 
and a termination instruction.
It is among other things shown that a function on bit strings whose
result is a bit string of length $1$ belongs to P/poly iff it can be 
computed by polynomial-length instruction sequences of this kind.
In~\cite{BB10a}, instruction sequences are considered which contain 
backward jump instructions in addition to the above-mentioned 
instructions.
It is among other things shown that a function on bit strings whose
result is a bit string of length $1$ belongs to PSPACE/poly iff it can 
be computed by polynomial-length instruction sequences of this latter 
kind.

It is known that $\textrm{NP} \subseteq \textrm{PSPACE/poly}$
(see e.g.~\cite{HS01a}).
Under the assumption that the reasonable complexity theoretic conjecture 
that $\textrm{NP} \not\subseteq \textrm{P/poly}$ (see e.g.~\cite{KL80a}) 
is right, it then follows that there exists a function on bit strings 
that can be computed by polynomial-length instruction sequences with 
backward jump instructions and cannot be computed by polynomial-length 
instruction sequences without backward jump instructions.
With this it remains among other things unanswered whether there exists 
a function on bit strings that can be computed by linear-length 
instruction sequences with backward jump instructions while it is 
commonly assumed that the function concerned cannot be computed by 
linear-length instruction sequences without backward jump instructions.
In this paper, we answer this question in the affirmative by means of a 
concrete example from binary arithmetic.

In~\cite{BM13b}, a description is given of instruction sequences of the 
kind used in~\cite{BM13a} that compute the hash function SHA-256 
according to the algorithm whose pseudo-code description serves as the 
definition of SHA-256 in the Secure Hash Standard~\cite{NIST12a}.
In~\cite{BM13c}, a description is given of instruction sequences of the 
kind used in~\cite{BM13a} that compute the function on bit strings that 
models the multiplication of natural numbers on their binary 
representation according to the Karatsuba multiplication 
algorithm~\cite{Kar95a,KO62a} and a description is given of instruction 
sequences of this kind that compute this function according to the 
standard multiplication algorithm, which is known as the long 
multiplication algorithm.
Thus, mathematically precise alternatives are provided to the natural 
language and pseudo-code descriptions of these algorithms found in the
literature on them.

In~\cite{BM13c}, the descriptions are further used to determine lower 
and upper estimates for the length of the representation in the binary 
number system of natural numbers at which the Karatsuba multiplication 
algorithm becomes more efficient than the long multiplication algorithm.
As expected, the function on bit strings that models the multiplication 
of natural numbers on their representation in the binary number system 
can be computed according to the long multiplication algorithm by 
quadratic-length instruction sequences without backward jump 
instructions.
Although it would not make the algorithm more efficient, it would be of 
practical value if this could be reduced to linear-length instruction 
sequences with backward jump instructions.
At first sight, this seems impossible unless provision is made for some 
form of indirect addressing for Boolean registers.
However, using a minor variant of the long multiplication algorithm, we 
will show in this paper that such a reduction is possible without making 
provision for some form of indirect addressing.
Riding off on an interesting side-issue, we will also sketch that even 
further reduction is possible if provision is made for some form of 
indirect addressing for Boolean registers.

It is customary that computing practitioners phrase their explanations 
of issues concerning programs from an empirical perspective such as 
the perspective that a program is in essence an instruction sequence.
An attempt to approach the semantics of programming languages from this 
perspective is made in~\cite{BL02a}.
The groundwork for the approach is an algebraic theory of single-pass
instruction sequences, called program algebra, and an algebraic theory
of mathematical objects that represent the behaviours produced by
instruction sequences under execution, called basic thread algebra.%
\footnote
{In~\cite{BL02a}, basic thread algebra is introduced under the name
 basic polarized process algebra.
}

As a continuation of this work on an approach to programming language
semantics, (a)~the notion of an instruction sequence was subjected to
systematic and precise analysis using the groundwork laid earlier and
(b)~selected issues relating to well-known subjects from the theory of
computation and the area of computer architecture were rigorously
investigated thinking in terms of instruction sequences
(see e.g.~\cite{BM12b}).
As in the work referred to above, the work presented in this paper is 
carried out in the setting of program algebra.
Different from usual in the work referred to above, but as in the work
presented in~\cite{BM13c,BM13b}, the accent is this time on a practical 
issues such as efficiency of algorithms and compactness of instruction 
sequences.

This paper is organized as follows.
First, we survey program algebra and the particular fragment and 
instantiation of it that is used in this paper (Section~\ref{sect-PGA}).
Next, we describe how we deal with $n$-bit words by means of Boolean 
registers (Section~\ref{sect-words}) and how we compute the basic 
operations on $n$-bit words that are used in the multiplication 
algorithms (Section~\ref{sect-opns-words}).
Then, we show that the function that models the multiplication of 
natural numbers on their representation in the binary number system can 
be computed according to a minor variant of the long multiplication 
algorithm by quadratic-length instruction sequences without backward 
jump instructions and by linear-length instruction sequences with 
backward jump instructions (Section~\ref{sect-lmul}).
After that, we discuss two points, concerning the halting problem and 
the concept of an algorithm, which were raised by the preceding material
(Sections~\ref{sect-hp} and~\ref{sect-alg}).
Following this, we sketch that for the algorithm under consideration 
further reduction is possible if provision is made for some form of 
indirect addressing for Boolean registers (Section~\ref{sect-ind-addr}).
Finally, we make some concluding remarks (Section~\ref{sect-concl}).

The preliminaries to the work presented in this paper are the same as 
the preliminaries to the work presented in~\cite{BM13c,BM13b}, which are 
in turn a selection from the preliminaries to the work presented 
in~\cite{BM13a}.
For this reason, there is some text overlap with those papers.
The preliminaries concern program algebra.
We only give a brief summary of program algebra.
A comprehensive introduction, including examples, can among other things
be found in~\cite{BM12b}.

\section{Program Algebra}
\label{sect-PGA}

In this section, we present a brief outline of \PGA\ (ProGram Algebra) 
and the particular fragment and instantiation of it that is used in 
the remainder of this paper.
A mathematically precise treatment can be found in~\cite{BM13a}.

The starting-point of \PGA\ is the simple and appealing perception
of a sequential program as a single-pass instruction sequence, i.e.\ a
finite or infinite sequence of instructions of which each instruction is
executed at most once and can be dropped after it has been executed or
jumped over.

It is assumed that a fixed but arbitrary set $\BInstr$ of
\emph{basic instructions} has been given.
The intuition is that the execution of a basic instruction may modify a 
state and produces a reply at its completion.
The possible replies are $\False$ and $\True$.
The actual reply is generally state-dependent.
Therefore, successive executions of the same basic instruction may
produce different replies.
The set $\BInstr$ is the basis for the set of instructions that may 
occur in the instruction sequences considered in \PGA.
The elements of the latter set are called \emph{primitive instructions}.
There are five kinds of primitive instructions, which are listed below:
\begin{itemize}
\item
for each $a \in \BInstr$, a \emph{plain basic instruction} $a$;
\item
for each $a \in \BInstr$, a \emph{positive test instruction} $\ptst{a}$;
\item
for each $a \in \BInstr$, a \emph{negative test instruction} $\ntst{a}$;
\item
for each $l \in \Nat$, a \emph{forward jump instruction} $\fjmp{l}$;
\item
a \emph{termination instruction} $\halt$.
\end{itemize}
We write $\PInstr$ for the set of all primitive instructions.

On execution of an instruction sequence, these primitive instructions
have the following effects:
\begin{itemize}
\item
the effect of a positive test instruction $\ptst{a}$ is that basic
instruction $a$ is executed and execution proceeds with the next
primitive instruction if $\True$ is produced and otherwise the next
primitive instruction is skipped and execution proceeds with the
primitive instruction following the skipped one --- if there is no
primitive instruction to proceed with,
inaction occurs;
\item
the effect of a negative test instruction $\ntst{a}$ is the same as
the effect of $\ptst{a}$, but with the role of the value produced
reversed;
\item
the effect of a plain basic instruction $a$ is the same as the effect
of $\ptst{a}$, but execution always proceeds as if $\True$ is produced;
\item
the effect of a forward jump instruction $\fjmp{l}$ is that execution
proceeds with the $l$th next primitive instruction of the instruction
sequence concerned --- if $l$ equals $0$ or there is no primitive
instruction to proceed with, inaction occurs;
\item
the effect of the termination instruction $\halt$ is that execution
terminates.
\end{itemize}

To build terms, \PGA\ has a constant for each primitive instruction and 
two operators. 
These operators are: the binary concatenation operator ${} \conc {}$ and 
the unary repetition operator ${}\rep$.
We use the notation $\Conc{i = 0}{n} P_i$, where $P_0,\ldots,P_n$ are 
\PGA\ terms, for the \PGA\ term $P_0 \conc \ldots \conc P_n$.
We also use the notation $P^n$. 
For each \PGA\ term $P$ and $n > 0$, $P^n$ is the \PGA\ term defined by 
induction on $n$ as follows: $P^1 = P$ and $P^{n+1} = P \conc P^n$.

The instruction sequences that concern us in the remainder of this paper 
are the finite ones, i.e.\ the ones that can be denoted by closed \PGA\ 
terms in which the repetition operator does not occur. 
Moreover, the basic instructions that concern us are instructions to set 
and get the content of Boolean registers.
More precisely, we take the set
\begin{ldispl}
\set{\inbr{i}.\getbr \where i \in \Natpos} \union
\set{\outbr{i}.\setbr{b} \where i \in \Natpos \Land b \in \Bool}
\\ \;\; {} \union
\set{\auxbr{i}.\getbr \where i \in \Natpos} \union
\set{\auxbr{i}.\setbr{b} \where i \in \Natpos \Land b \in \Bool} 
\end{ldispl}%
as the set $\BInstr$ of basic instructions.

Each basic instruction consists of two parts separated by a dot.
The part on the left-hand side of the dot plays the role of the name of 
a Boolean register and the part on the right-hand side of the dot plays 
the role of a command to be carried out on the named Boolean register.
For each $i \in \Natpos$:
\begin{itemize}
\item
$\inbr{i}$ serves as the name of the Boolean register that is used as 
$i$th input register in instruction sequences;
\item
$\outbr{i}$ serves as the name of the Boolean register that is used as
$i$th output register in instruction sequences;
\item
$\auxbr{i}$ serves as the name of the Boolean register that is used as 
$i$th auxiliary register in instruction sequences.
\end{itemize}
On execution of a basic instruction, the commands have the following 
effects:
\begin{itemize}
\item
the effect of $\getbr$ is that nothing changes and the reply is the 
content of the named Boolean register;
\item
the effect of $\setbr{\False}$ is that the content of the named Boolean 
register becomes $\False$ and the reply is $\False$;
\item
the effect of $\setbr{\True}$ is that the content of the named Boolean 
register becomes $\True$ and the reply is $\True$.
\end{itemize}

We are also interested in the extension of \PGA\ with, for each 
$l \in \Nat$, a \emph{backward jump instruction} $\bjmp{l}$ as
additional primitive instruction.
On execution of an instruction sequence, the effect of a backward jump 
instruction $\bjmp{l}$ is that execution proceeds with the $l$th 
previous primitive instruction of the instruction sequence concerned --- 
if $l$ equals $0$ or there is no primitive instruction to proceed with, 
inaction occurs.
We write \PGAbj\ for \PGA\ with these additional primitive instructions.

Regarding the behaviours produced by finite instruction sequences with 
backward jump instructions under execution, we refer to the treatment 
of C, which is a variant of \PGA, in~\cite{BP09a}.
The fragment of \PGAbj\ without the repetition operator coincides with 
the fragment of C without backward instructions other than backward jump 
instructions.

Let $n,m \in \Nat$, let $\funct{f}{\set{0,1}^n}{\set{0,1}^m}$, and let
$X$ be a finite instruction sequence that can be denoted by a closed 
\PGA\ or \PGAbj\ term in the case that $\BInstr$ is taken as specified 
above.
Then $X$ \emph{computes} $f$ if there exists a $k \in \Nat$ such that, 
for all $b_1,\ldots,b_n \in \Bool$, if $X$ is executed in an environment 
with $n$ input registers, $m$ output registers, and $k$ auxiliary 
registers, the content of the input registers with names 
$\inbr{1},\ldots,\inbr{n}$ are $b_1,\ldots,b_n$ when execution starts, 
and the content of the output registers with names 
$\outbr{1},\ldots,\outbr{m}$ are $b'_1,\ldots,b'_m$ when execution 
terminates, then $f(b_1,\ldots,b_n) = b'_1,\ldots,b'_m$.

Let $\funct{f}{\seqof{\set{0,1}}}{\seqof{\set{0,1}}}$ be such that for
all $\beta,\beta' \in \seqof{\set{0,1}}$, $\len(\beta) = \len(\beta')$
implies $\len(f(\beta)) = \len(f(\beta'))$, and 
let $F \subseteq \set{g \where \funct{g}{\Nat}{\Nat}}$.
Then $f$ \emph{can be computed by $F$-length instruction sequences} if 
there exists a $g \in F$ such that, for all $n \in \Nat$, there exists a 
finite instruction sequence $X$ that can be denoted by a closed \PGA\ or 
\PGAbj\ term such that $X$ computes the restriction of $f$ to 
$\set{0,1}^n$ and $\len(X) \leq g(n)$.
We write \emph{polynomial-length} instead of $F$-length if $F$ is the 
set of all polynomial functions $\funct{g}{\Nat}{\Nat}$.
The phrases \emph{quadratic-length} and \emph{linear-length} are used 
similarly.

\section{Dealing with $n$-Bit Words}
\label{sect-words}

This section is concerned with dealing with bit strings of length $n$ 
by means of Boolean registers.
It contains definitions which facilitate the description of instruction 
sequences that compute the function on bit strings that models the 
multiplication of natural numbers on their representation in the binary 
number system according to the long multiplication algorithm or a minor 
variant thereof.
In the sequel, bit strings of length $n$ will mostly be called 
\emph{$n$-bit words}.
The prefix ``$n$-bit'' is left out if $n$ is irrelevant or clear from
the context.

Let $\kappa{:}i$ 
($\kappa \in \set{\mathsf{in},\mathsf{out},\mathsf{aux}}$, 
 $i \in \Natpos$) be the name of a Boolean register.
Then $\kappa$ and $i$ are called the \emph{kind} and \emph{number} of 
the Boolean register.
Successive Boolean registers are Boolean registers of the same kind with
successive numbers.
Words are stored by means of Boolean registers such that the successive 
bits of a stored word are the content of successive Boolean registers.

Henceforth, the name of a Boolean register will mostly be used to refer 
to the Boolean register in which the least significant bit of a word is 
stored.
Let $\kappa{:}i$ and $\kappa'{:}i'$ be the names of Boolean registers 
and let $n \in \Natpos$.
Then we say that $\kappa{:}i$ and $\kappa'{:}i'$ \emph{lead to partially 
coinciding $n$-bit words} if $k = k'$ and $|i - i'| < n$.

The words that represent the two natural numbers whose product is to be 
computed are stored in advance of the whole computation in input 
registers, starting with the input register with number $1$.
It is convenient to have available, for each $n > 0$, the names 
$I_1^{(n)}$ and $I_2^{(n)}$ for the input registers in which the least 
significant bit of these words are stored.
The word that represents the product is stored before the end of the 
whole computation in output registers, starting with the output register 
with number $1$. 
It is convenient to have available, for each $n > 0$, the name 
$O^{(n)}$ for the output register in which the least significant bit of 
this word is stored.

A number of words that represent intermediate values computed are 
temporarily stored during the whole computation in auxiliary registers, 
starting with the auxiliary register with number $1$.
It is convenient to have available, for each $n > 0$, names $T_1^{(n)}$, 
$T_2^{(n)}$, \ldots\ for auxiliary registers in which the least 
significant bit of these words are stored.
Moreover, it is convenient to have available the name $c$ for the 
auxiliary register that contains the carry bit that is repeatedly stored 
when computing the function on bit strings that models the addition of 
natural numbers on their representation in the binary number system.

Therefore, we define for each $n > 0$ and $i > 0$: 
\begin{ldispl}
\begin{asceqns}
I_1^{(n)} & \deq & \inbr{1},   \\
I_2^{(n)} & \deq & \inbr{k}
          & \mathrm{where}\; k = n + 1, \\
O^{(n)}   & \deq & \outbr{1},  \\
T_i^{(n)} & \deq & \auxbr{k}
          & \mathrm{where}\; k = 2 \mul n \mul (i - 1) + 2, \\
c         & \deq & \auxbr{1}.
\end{asceqns}
\end{ldispl}%
For each $n > 0$, $I_1^{(n)}$, $I_2^{(n)}$, $O^{(n)}$, $T_1^{(n)}$, 
$T_2^{(n)}$, $T_3^{(n)}$, and $T_4^{(n)}$ are the names 
that will be used in Section~\ref{sect-lmul} to define instruction 
sequences that compute the function on bit strings of length $n$ that 
models the multiplication of two natural numbers less than $2^n$ on 
their representation in the binary number system.
Moreover, we will write $I_i^{(n)}[j]$ ($0 \leq j < n$) for $\inbr{k}$ 
where $k = (i - 1) \mul n + j + 1$ and 
$T_i^{(n)}[j]$ ($0 \leq j < 2 \mul n$) for $\auxbr{k}$ 
where $k = 2 \mul (i - 1) \mul n + j + 2$.

\section{Computing Operations on $n$-Bit Words}
\label{sect-opns-words}

This section is concerned with computing operations on bit strings of 
length $n$.
It contains definitions which facilitate the description of instruction 
sequences that compute the function on bit strings that models the 
multiplication of natural numbers on their representation in the binary 
number system according to the long multiplication algorithm or a minor 
variant thereof.

Henceforth, we will write $\beta \beta'$, where $\beta$ and $\beta'$ are 
bit strings, for the concatenation of $\beta$ and $\beta'$.
In other words, we will use juxtaposition for concatenation.
Moreover, we will use the bit string notation $b^n$.
For $n > 0$, the bit string $b^n$, where $b \in \set{0,1}$, is defined 
by induction on $n$ as follows: $b^1 = b$ and $b^{n+1} = b\,b^n$.

The basic operations on words that are relevant to the different 
multiplica\-tion algorithms are test on nonzero, decrement by one, 
shift left $m$ positions \linebreak[2] ($0 < m < n$), shift right $m$ 
positions ($0 < m < n$), and addition on $n$-bit words ($n > 0$).
For these operations, we define parameterized instruction sequences 
computing them in case the parameters are properly instantiated (see 
below):
\begin{ldispl}
\TSTNZ{n}(\srcbr{k}) \deq {}
\\ \quad
\Conc{i = 0}{n-1} 
 (\ptst{\srcbr{k{+}i}.\getbr} \conc \fjmp{2} \conc \fjmp{3}) \conc
  \fjmp{1}\;,
\eqnsep
\DEC{n}(\srcbr{k},\dstbr{l}) \deq {}
\\ \quad
\Conc{i = 0}{n-1} 
 (\ntst{\srcbr{k{+}i}.\getbr} \conc \fjmp{3} \conc
  \dstbr{l{+}i}.\setbr{\False} \conc \fjmp{5} \conc
  \dstbr{l{+}i}.\setbr{\True}) \conc
\fjmp{1} \conc \fjmp{1} \conc \fjmp{1}\;,
\eqnsep
\SHL{n}{m}(\srcbr{k},\dstbr{l}) \deq {}
\\ \quad
\Conc{i = 0}{n-1-m} 
 (\ptst{\srcbr{k{+}n{-}1{-}m{-}i}.\getbr} \conc \fjmp{2} \conc
  \ptst{\dstbr{l{+}n{-}1{-}i}.\setbr{\False}} \conc
  \dstbr{l{+}n{-}1{-}i}.\setbr{\True}) \conc {}
\\[.5ex] \quad
\Conc{i = 0}{m-1} (\dstbr{l{+}m{-}1{-}i}.\setbr{\False})\;,
\eqnsep
\SHR{n}{m}(\srcbr{k},\dstbr{l}) \deq {}
\\ \quad
\Conc{i = 0}{n-1-m} 
 (\ptst{\srcbr{k{+}m{+}i}.\getbr} \conc \fjmp{2} \conc
  \ptst{\dstbr{l{+}i}.\setbr{\False}} \conc
  \dstbr{l{+}i}.\setbr{\True}) \conc {}
\\[.5ex] \quad
\Conc{i = 0}{m-1} (\dstbr{l{+}n{-}m{+}i}.\setbr{\False})\;,
\eqnsep
\ADD{n}(\srcbri{1}{k_1},\srcbri{2}{k_2},\dstbr{l}) \deq {} 
\\ \quad
c.\setbr{\False} \conc {}
\\ \quad
\Conc{i = 0}{n-1} 
 (\ptst{\srcbri{1}{k_1{+}i}.\getbr} \conc \fjmp{4}  \conc 
  \ptst{\srcbri{2}{k_2{+}i}.\getbr} \conc \fjmp{7} \conc \fjmp{9} \conc
  \ptst{\srcbri{2}{k_2{+}i}.\getbr} \conc \fjmp{10} \conc {}
\\ \quad \phantom{\Conc{i = 0}{n-1} (}
  \ptst{c.\getbr}               \conc \fjmp{10} \conc \fjmp{16} \conc 
  \ptst{c.\getbr}               \conc \fjmp{7} \conc \fjmp{13} \conc 
  \ptst{c.\getbr}               \conc \fjmp{11} \conc \fjmp{9} \conc 
  \ptst{c.\getbr}               \conc \fjmp{4} \conc {}
\\ \quad \phantom{\Conc{i = 0}{n-1} (}  
  \dstbr{l{+}i}.\setbr{\False} \conc c.\setbr{\True} \conc
  \fjmp{6} \conc
  \dstbr{l{+}i}.\setbr{\True}  \conc c.\setbr{\True} \conc 
  \fjmp{3} \conc {}
\\ \quad \phantom{\Conc{i = 0}{n-1} (}
  \ptst{\dstbr{l{+}i}.\setbr{\False}} \conc 
  \dstbr{l{+}i}.\setbr{\True})\;,
\end{ldispl}%
where 
$s,s_1,s_2$ range over $\set{\mathsf{in},\mathsf{aux}}$, 
$d$ ranges over $\set{\mathsf{aux},\mathsf{out}}$, and
$k,k_1,k_2,l$ range over $\Natpos$.
For each of these parameterized instruction sequences except the first 
one, all but the last parameter correspond to the operands of the 
operation concerned and the last parameter corresponds to the result of 
the operation concerned.
The intended operations are computed provided that the instantiation of 
the last parameter and the instantiation of none of the other parameters 
lead to partially coinciding $n$-bit words.
In this paper, this condition will always be satisfied.
No result is stored on execution of $\TSTNZ{n}$.
Instead, the first primitive instruction following $\TSTNZ{n}$ is 
skipped if the test on nonzero fails.

Transferring $n$-bit words ($n > 0$) is also relevant to multiplication 
algorithms.
For this, we define parameterized instruction sequences as well.
By one the successive bits in a constant $n$-bit word become the content 
of $n$ successive Boolean registers and by the other the successive bits 
in a $n$-bit word that are the content of $n$ successive Boolean 
registers become the content of $n$ other successive Boolean registers:
\begin{ldispl}
\SET{n}(b_0 \ldots b_{n-1},\dstbr{l}) \deq 
\Conc{i = 0}{n-1} (\dstbr{l{+}i}.\setbr{b_i})\;,
\eqnsep
\MOV{n}(\srcbr{k},\dstbr{l}) \deq 
\Conc{i = 0}{n-1} 
 (\ptst{\srcbr{k{+}i}.\getbr} \conc \fjmp{2} \conc
  \ptst{\dstbr{l{+}i}.\setbr{\False}} \conc 
  \dstbr{l{+}i}.\setbr{\True})\;,
\end{ldispl}%
\sloppy
where 
$b_0,\ldots,b_{n-1}$ range over $\set{\False,\True}$,
$s$ ranges over $\set{\mathsf{in},\mathsf{aux}}$, 
$d$ ranges over $\set{\mathsf{aux},\mathsf{out}}$, and
$k,l$ range over $\Natpos$.
In the case of $\MOV{n}$, the intended transfer is performed provided 
that the instantiation of the last parameter and the instantiation of 
the first parameter do not lead to partially coinciding $n$-bit words.
In this paper, this condition will always be satisfied.

For convenience's sake, we define a special case of the 
parameterized instruction sequences for transferring $n$-bit words 
($0 < m < n$):
\begin{ldispl}
\ZPAD{n}{m}(\dstbr{l}) \deq
\SET{n-m}(0^{n-m},\dstbr{l{+}m})\;,
\end{ldispl}%
where 
$d$ ranges over $\set{\mathsf{aux},\mathsf{out}}$ and
$l$ range over $\Natpos$.
$\ZPAD{n}{m}$ is meant for turning a stored $m$-bit word into a stored 
$n$-bit word by zero padding.

The calculation of the lengths of the parameterized instruction 
sequences defined above is a matter of simple additions and 
multiplications. 
The lengths of these instruction sequences are as follows:
\begin{ldispl}
\len(\TSTNZ{n}(\srcbr{k})) = 3 \mul n + 1\;, \\
\len(\DEC{n}(\srcbr{k},\dstbr{l})) = 5 \mul n + 3\;, \\
\len(\SHL{n}{m}(\srcbr{k},\dstbr{l})) = 4 \mul n - 3 \mul m\;, \\
\len(\SHR{n}{m}(\srcbr{k},\dstbr{l})) = 4 \mul n - 3 \mul m\;, \\
\len(\ADD{n}(\srcbri{1}{k_1},\srcbri{2}{k_2},\dstbr{l})) = 
 26 \mul n + 1\;, 
\eqnsep 
\len(\SET{n}(b_0 \ldots b_{n-1},\dstbr{l})) = n\;, \\
\len(\MOV{n}(\srcbr{k},\dstbr{l})) = 4 \mul n\;,  \\
\len(\ZPAD{n}{m}(\dstbr{l})) = n - m\;.
\end{ldispl}%

Note that the instruction sequences defined in this section do compute 
the intended operations in case of fully coinciding $n$-bit words.
Slightly shorter instruction sequences are defined for addition on 
$n$-bit words and transfer of a stored $n$-bit word in~\cite{BM13b}, but 
those instruction sequences do not compute the intended operations in 
case of fully coinciding $n$-bit words. 

\section{Long Multiplication and Backward Jump Instructions}
\label{sect-lmul}

This section shows that the function on bit strings that models the 
multiplication of natural numbers on their representation in the binary 
number system can be computed according to a minor variant of the long 
multiplication algorithm by quadratic-length instruction sequences 
without backward jump instructions and by linear-length instruction 
sequences with backward jump instructions.

We begin with defining instruction sequences without backward jump 
instructions that compute this function according to the long 
multiplication algorithm.
The additions are done on the fly and the shifts are restricted to one 
position by shifting the result of all preceding shifts.

We uniformly define instruction sequences $\LMULi{n}$ ($n > 0$) by 
\begin{ldispl}
\MOV{n}(I_1^{(n)},T_1^{(n)}) \conc \ZPAD{2n}{n}(T_1^{(n)}) \conc
\SET{2n}(0^{2n},T_2^{(n)}) \conc {}
\\ 
\Conc{i = 0}{n-1} 
 \bigl(\ntst{I_2^{(n)}[i].\getbr} \conc \fjmp{l_i} \conc   
  \ADD{n+i+1}(T_1^{(n)},T_2^{(n)},T_2^{(n)}) \conc 
  \SHL{n+i+1}{1}(T_1^{(n)},T_1^{(n)})
 \bigr) \conc {}
\\ 
\MOV{2n}(T_2^{(n)},O^{(n)}) \conc \halt\;,
\\[1.1ex]
\mbox{where} 
\\[1.1ex]
l_i = \len(\ADD{n+i+1}(T_1^{(n)},T_2^{(n)},T_2^{(n)})) + 1 =
      26 \mul n + 26 \mul i + 28
      \;\; (0 \leq i \leq n - 1).
\end{ldispl}%
Using the property that $\sum_{i = 0}^k i = (k \mul (k + 1)) / 2$, we
obtain by simple calculations that 
\begin{ldispl}
\len(\LMULi{n}) = 45 \mul n^2 + 30 \mul n + 1\;.
\end{ldispl}%
This means that the function on bit strings that models the 
multiplication of natural numbers on their representation in the binary 
number system can be computed by quadratic-length instruction sequences 
without backward jump instructions if it is computed according to the 
long multiplication algorithm.

For each bit of the representation of the multiplier, $\LMULi{n}$ 
contains a different instruction sequence.
This seems to exclude the use of backward jump instructions to obtain 
linear-length instruction sequences, unless provision is made for some
form of indirect addressing for Boolean registers.
However, there exists a minor variant of the long multiplication 
algorithm that makes it possible to have the same instruction sequence 
for each bit of the representation of the multiplier.
From the least significant bit of the representation of the multiplier
onwards, the algorithm concerned shifts the representation of the 
multiplier one position to the right after it has dealt with a bit.
In this way, the next bit remains the least significant one throughout.

We proceed with defining instruction sequences without backward jump 
instructions that compute the function on bit strings that models the 
multiplication of natural numbers on their representation in the binary 
number system according to this minor variant of the long multiplication 
algorithm.

We uniformly define instruction sequences $\LMULii{n}$ ($n > 0$) by 
\begin{ldispl}
\MOV{n}(I_1^{(n)},T_1^{(n)}) \conc \ZPAD{2n}{n}(T_1^{(n)}) \conc
\MOV{n}(I_2^{(n)},T_2^{(n)}) \conc \SET{2n}(0^{2n},T_3^{(n)}) \conc {}
\\ 
 \bigl(
  \ntst{T_2^{(n)}[0].\getbr} \conc \fjmp{l} \conc   
  \ADD{2n}(T_1^{(n)},T_3^{(n)},T_3^{(n)}) \conc {}
\\ \phantom{\bigl(}
  \SHL{2n}{1}(T_1^{(n)},T_1^{(n)}) \conc
  \SHR{n}{1}(T_2^{(n)},T_2^{(n)})
 \bigr)^n \conc {}
\\ 
\MOV{2n}(T_3^{(n)},O^{(n)}) \conc \halt\;,
\\[1.1ex]
\mbox{where} 
\\[1.1ex]
l = \len(\ADD{2n}(T_1^{(n)},T_3^{(n)},T_3^{(n)})) + 1 = 52 \mul n + 2\;.
\end{ldispl}%
We obtain by simple calculations that 
\begin{ldispl}
\len(\LMULii{n}) = 64 \mul n^2 + 16 \mul n + 1\;.
\end{ldispl}%
This means that the function on bit strings that models the 
multiplication of natural numbers on their representation in the binary 
number system can still be computed by quadratic-length instruction 
sequences without backward jump instructions if it is computed according 
to the minor variant of the long multiplication algorithm.
Moreover, we have that $\len(\LMULii{n}) > \len(\LMULi{n})$ for all 
$n > 0$.

For each bit of the representation of the multiplier, $\LMULii{n}$ 
contains the same instruction sequence.
That is, it contains $n$ duplicates of the same instruction sequence.
This duplication can be eliminated by implementing a for loop by means
of a backward jump instruction.

We proceed with defining instruction sequences with backward jump 
instructions that compute the function on bit strings that models the 
multiplication of natural numbers on their representation in the binary 
number system according to the minor variant of the long multiplication 
algorithm.
In the definition to come, we write $\overline{n}$ for the shortest 
representation of the natural number $n$ in the binary number system.

We uniformly define instruction sequences $\LMULiii{n}$ ($n > 0$) by 
\begin{ldispl}
\MOV{n}(I_1^{(n)},T_1^{(n)}) \conc \ZPAD{2n}{n}(T_1^{(n)}) \conc
\MOV{n}(I_2^{(n)},T_2^{(n)}) \conc \SET{2n}(0^{2n},T_3^{(n)}) \conc {}
\\
\SET{\floor{\log_2(n)}+1}(\overline{n},T_4^{(n)}) \conc {}
\\ 
\ntst{T_2^{(n)}[0].\getbr} \conc \fjmp{l_1} \conc   
\ADD{2n}(T_1^{(n)},T_3^{(n)},T_3^{(n)}) \conc {} 
\\
\SHL{2n}{1}(T_1^{(n)},T_1^{(n)}) \conc
\SHR{n}{1}(T_2^{(n)},T_2^{(n)}) \conc {}
\\
\DEC{\floor{\log_2(n)}+1}(T_4^{(n)},T_4^{(n)}) \conc
\TSTNZ{\floor{\log_2(n)}+1}(T_4^{(n)}) \conc \bjmp{l_2} \conc {}
\\
\MOV{2n}(T_3^{(n)},O^{(n)}) \conc \halt\;,
\end{ldispl}%
where
\begin{ldispl}
l_1 = \len(\ADD{2n}(T_1^{(n)},T_3^{(n)},T_3^{(n)})) + 1 = 
      52 \mul n + 2\;,
\\
l_2 = \len(\ntst{T_2^{(n)}[0].\getbr} \conc \ldots \conc 
           \TSTNZ{\floor{\log_2(n)}+1}(T_4^{(n)})) = 
       64 \mul n + 9 \mul \floor{\log_2(n)} + 11\;.
\end{ldispl}%
We obtain by simple calculations that 
\begin{ldispl}
\len(\LMULiii{n}) = 83 \mul n + 9 \mul \floor{\log_2(n)} + 12\;.
\end{ldispl}%
This means that the function on bit strings that models the 
multiplication of natural numbers on their representation in the binary 
number system can be computed by linear-length instruction sequences 
with backward jump instructions if it is computed  according to the 
minor variant of the long multiplication algorithm.
Moreover, we have that $\len(\LMULiii{n}) < \len(\LMULi{n})$ for all 
$n > 1$.

\section{Long Multiplication and the Halting Problem}
\label{sect-hp}

In this section, a point concerning the halting problem is discussed 
which was raised by the material in Section~\ref{sect-lmul}, but for 
which space could not be found there.

Turing's result regarding the undecidability of the halting problem 
(see e.g.~\cite{Tur37a}) is a result about Turing machines.
In~\cite{BM09k}, we consider it as a result about programs rather than 
machines, taking instruction sequences as programs.
The instruction sequences concerned are essentially the finite 
instruction sequences that can be denoted by closed \PGAbj\ terms.
Unlike in the current paper, the basic instructions are not fixed, but 
their effects are restricted to the manipulation of something that can 
be understood as the content of the tape of a Turing machine with a 
specific tape alphabet, together with the position of the tape head.
Different choices of basic instructions give rise to different halting 
problem instances and one of these instances is essentially the same as 
the halting problem for Turing machines.
Because of their orientation to Turing machines, we consider all 
instances treated in~\cite{BM09k} theoretical halting problem instances.

All halting problem instances would evaporate if the instruction 
sequences concerned would be restricted to the ones without backward 
jump instructions.
This is irrespective of whether the effects of the basic instructions 
have anything to do with the manipulation of a Turing machine tape.
In the case that we have basic instructions to set and get the content 
of Boolean registers, instruction sequences without backward jump 
instructions are sufficient to compute all functions 
$\funct{f}{\set{0,1}^n}{\set{0,1}^m}$ ($n,m \in \Nat$).
This raises the question whether there exists a good reason for not 
abandoning backward jump instructions altogether in such cases.
The function that models the multiplication of natural numbers on their 
representation in the binary number system offers a good reason: the 
length of the instruction sequences that compute it according to the
long multiplication algorithm can be reduced significantly by the use of
backward jump instructions, even more than by going over to one of the
multiplication algorithms that are known to yield shorter instruction
sequences without backward jump instructions than the long 
multiplication algorithm such as for example the Karatsuba 
multiplication algorithm (see e.g.~\cite{BM13c}).

Thus, the instruction sequences $\LMULii{n}$ and the instruction 
sequences $\LMULiii{n}$ form a hard witness of the inevitable 
existence of a halting problem in the practice of imperative 
programming, where programs must have manageable size.
Because of its orientation to actual programming, we consider the 
halting problem for the instruction sequences with forward and backward 
jump instructions, and with only basic instructions to set and get the 
content of Boolean registers, a practical halting problem.
It is unknown to us whether there is a connection between the 
solvability or unsolvability of the halting problem for these 
instruction sequences and some form of diagonal argument.
It is easy to prove that this halting problem is both NP-hard and 
coNP-hard.
We do not know whether stronger lower bounds for its complexity can be 
found in the literature.
An extensive search for such lower bounds and other result concerning 
this halting problem or a similar halting problem has been unsuccessful.

\section{Long Multiplication and the Concept of an Algorithm}
\label{sect-alg}

In this section, another point is discussed which was raised by the 
material in Section~\ref{sect-lmul}.
This point concerns the concept of an algorithm.

At the end of Section~\ref{sect-lmul}, we implicitly state that the
instruction sequences $\LMULii{n}$ and the instruction sequences 
$\LMULiii{n}$ realize the same algorithm.
We have asked ourselves the question why this is an acceptable statement 
and what this says about the definition of an algorithm.
We consider it an acceptable statement because all the different views 
on what characterizes an algorithm lead to the conclusion that we have 
to do here with different realizations of the same algorithm.
We cannot prove this due to the absence of a generally accepted 
mathematically precise definition of the concept of an algorithm.
The cause of this absence seems to be the general acceptance of the 
exact mathematical concept of a Turing machine and equivalent 
mathematical concepts as adequate replacements of the intuitive concept 
of an algorithm.

Unfortunately, Turing machines are quite remote from anything related to 
actual programming.
Moreover, we can construct at least two different Turing machines for 
the one algorithm realized by both the instruction sequences 
$\LMULii{n}$ and the instruction sequences $\LMULiii{n}$: one without a 
counterpart of a for loop and one with a counterpart of a for loop.
So Turing machines do not enforce a level of abstraction that is 
sufficient for algorithms.
Therefore, we doubt whether the mathematical concept of a Turing machine 
is an adequate replacement of the intuitive concept of an algorithm.
This means that we consider a generally accepted mathematically precise 
definition of the concept of an algorithm still desirable.
Below, we outline a possible avenue to such a\linebreak[2] definition.

We restrict ourselves to algorithms for computing functions on bit 
strings.
This has the advantage that data representation is hardly an issue in 
the realizations of algorithms.
Moreover, we adopt the common practice among mathematicians to treat the
length of the input of an algorithm as a parameter of the algorithm.
In the perspective that a program is in essence an instruction sequence, 
taking into account the experience gained in this paper with realizing 
algorithms by instruction sequences, we consider the following to be a 
first approximation of a mathematically precise definition of the 
concept of an algorithm: ``an algorithm is a mapping from the set of 
natural numbers to the set of equivalence classes of the instruction 
sequences with backward jumps used in this paper with respect to an 
appropriate equivalence relation''.
The underlying idea is that for each algorithm, for each $n$, there is
a class of algorithmically equivalent instruction sequences that realize
the algorithm for that $n$.
This idea refines an idea that was already put forward by Milner in 1971 
(see~\cite{Mil71a}).

What exactly should be considered algorithmically equivalent instruction
sequences is a matter of further study.
Some requirements for algorithmic equivalence are:
\begin{itemize}
\item
each instruction sequence is algorithmically equivalent to each 
instruction sequence that produces the same behaviour;
\item
each instruction sequence is algorithmically equivalent to the 
instruction sequence obtained from it by consistently exchanging $0$ and 
$1$;
\item
each instruction sequence is algorithmically equivalent to each 
instruction sequence obtained from it by renumbering the auxiliary 
Boolean registers~used;
\item
each instruction sequence is algorithmically equivalent to each
instruction sequence obtained from it by transposing basic instructions 
that have no influence on each other;
\item
each instruction sequence is algorithmically equivalent to each
instruction sequence obtained from it by replacing subsequences that are 
the result of the concatenation of an instruction sequence a number of 
times with itself by an implementation of a for loop of which it is the 
unwinding.
\end{itemize}
Of course, there is a possibility that additional requirements are 
necessary.
Note that $\LMULii{n}$ and $\LMULiii{n}$ are algorithmically equivalent
according to the last-mentioned requirement.
It is mainly this requirement that makes it difficult to give an exact 
mathematical definition of an algorithmic equivalence relation 
satisfying the above-mentioned requirements.
We further remark that it is not clear to us whether such a definition 
is relevant at all if the conceivable viewpoint is taken that there may
be different degrees to which an instruction sequence realizes an 
algorithm.

Above, we have restricted ourselves to algorithms for computing 
functions on bit strings.
We could restrict ourselves further to algorithms for computing 
projective functions on bit strings, i.e.\ functions on bit strings for 
which the restriction to bit strings of any given length can handle each 
restriction to bit strings of a shorter length if sufficiently many 
leading zeros are added (see~\cite{BM13a}).
This means that an instruction sequence that computes the restriction of
such a function to bit strings of a certain length can also be used to 
compute the restriction of the function concerned to bit strings up to 
that length.
The projective functions on bit strings include all functions that model 
operations on natural numbers on their representation in the binary 
number system.

\section{Further Reduction of Instruction Sequence Length}
\label{sect-ind-addr}

In Section~\ref{sect-lmul}, it is demonstrated that the function on bit 
strings that models the multiplication of natural numbers on their 
representation in the binary number system can be computed according to 
a minor variant of the long multiplication algorithm by linear-length 
instruction sequences with backward jump instructions.
The alteration of the long multiplication algorithm seems inescapable in 
this case unless provision is made for some form of indirect addressing 
for Boolean registers.
If such a provision is made, however, the function concerned cannot 
only be computed according to the unaltered long multiplication 
algorithm, but also by logarithmic-length instruction sequences with 
backward jump instructions.

The expression defining $\LMULi{n}$ contains a subexpression of the form 
$\Conc{i=0}{n-1} P_i$.
It is easy to see that, if provision is made for some form of indirect 
addressing for Boolean registers, there exists an instruction sequence 
$P$ such that this subexpression can be replaced by $P^n$.
The duplication of $P$ can then be eliminated by implementing a for loop 
like in $\LMULiii{n}$.
Because these remarks apply to $\SHL{n}{m}$, $\ADD{n}$, $\SET{n}$, 
$\MOV{n}$, and $\ZPAD{n}{m}$ as well, indirect addressing of Boolean 
registers makes logarithmic-length instruction sequences possible.

Some of the for loops that has to be implemented to obtain 
logarithmic-length instruction sequences require an increasing loop 
counter.
Instead of instruction sequences for test on nonzero ($\TSTNZ{n}$) and 
decrement by one ($\DEC{n}$), instruction sequences for test on not 
equal to $n$ and increment by one are needed to implement such a for 
loop.
An instruction sequence for setting the loop counter to its initial 
value is needed as well to implement a for loop.
To obtain logarithmic-length instruction sequences, it is sufficient to 
use $\SET{n}$ as defined in Section~\ref{sect-lmul} for this purpose. 

The form of indirect addressing known as indexed addressing in the area 
of computer architecture is most appropriate for the algorithm under 
consideration.
In the case of indexed addressing of a Boolean register, its number is
obtained by adding the number whose representation in the binary number
system is formed by the contents of specified successive Boolean 
registers to a specified number.
In the area of computer architecture, the latter number is usually
called the base address and the former number is usually called the 
index.
In the case of direct addressing, we use an expression of the form 
$\kappa{:}i$, where 
$\kappa \in \set{\mathsf{in},\mathsf{out},\mathsf{aux}}$ and 
$i \in \Natpos$, on the left-hand side of the dot in basic instructions 
to refer to the Boolean register of kind $\kappa$ whose number is $i$.
In the case of indexed addressing, we could use an expression of the 
form $\kappa{:}i(\auxbr{j}{:}l)$, where 
$\kappa \in \set{\mathsf{in},\mathsf{out},\mathsf{aux}}$ and 
$i,j,l \in \Natpos$, on the left-hand side of the dot in basic 
instructions to refer to the Boolean register of kind $\kappa$ whose 
number is the sum of $i$ and the number represented by the contents of 
the $l$ successive Boolean registers of kind $\mathsf{aux}$ of which the 
first one has number $j$. 

If $\LMULi{n}$ is adapted as outlined above, the length of the adapted 
instruction sequence is 
$c \mul \floor{\log_2(n)} + c' \mul \floor{\log_2(2n-1)} + c''$, where
$c$, $c'$, and $c''$ are constants greater than zero.
With indexed addressing, it is straightforward to obtain an instruction 
sequence $\LMULiv{n}$ such that $c < 100$, $c' < 10$, and $c'' < 250$.
This means that the function on bit strings that models the 
multiplication of natural numbers on their representation in the binary 
number system can be computed by logarithmic-length instruction 
sequences with backward jump instructions according to the long 
multiplication algorithm if provision is made for some form of indirect 
addressing for Boolean registers.
Moreover, we have that $\len(\LMULiv{n}) < \len(\LMULiii{n})$ for all 
$n > 5$.
This reduction of instruction sequence length is obtained by 
instructions that provide for backward jumping and a form of indirect 
addressing.
It is an open question whether it can be reduced further with 
instructions that provide for additional facilities.

\section{Concluding Remarks}
\label{sect-concl}

We have demonstrated that, in the case that the other instructions are 
only instructions to set and get the content of Boolean registers, 
forward jump instructions, and a termination instruction, the function 
that models the multiplication of natural numbers on their 
representation in the binary number system can be computed according to 
a minor variant of the long multiplication algorithm by quadratic-length 
instruction sequences without backward jump instructions and by 
linear-length instruction sequences with backward jump instructions.
Be aware that we have not shown that this function cannot be computed by
linear-length instruction sequences without backward jump instructions.
However, the scientific literature on multiplication algorithms (see 
e.g.~\cite{Fur09a,KO62a,SS71a,Too63a}) indicates that it is likely 
that it cannot be computed by linear-length instruction sequences 
without backward jump instructions.

We have also gone into the observations that the demonstration provides
a hard witness of the inevitable existence of a halting problem in the 
practice of imperative programming and that it makes manifest the lack 
of a definition of the concept of an algorithm that makes it possible to
prove whether two instruction sequences realize the same algorithm.

The viewpoints on what is an algorithm are diverse in character.
Milner's idea that algorithms are equivalence classes of programs can 
also be found in~\cite{Yan11a}.
A rather strange twist is that constructions of primitive recursive 
functions are considered to be programs.
In~\cite{Mos01b}, algorithms are viewed as isomorphism classes of tuples 
of recursive functionals that can be defined by repeated application of 
certain schemes.
In~\cite{BC82a}, which is concerned with algorithms on Kahn-Plotkin's 
concrete data structures, algorithms are viewed as pairs of a function 
and a computation strategy that resolves choices between possible ways 
of computing the function.
In~\cite{Gur00a}, an algorithm is defined as an object that satisfy 
certain postulates.
According to this definition, Gurevich's abstract state machines capture
algorithms.
In~\cite{KU58a}, it is claimed that the only algorithms are those 
realized by Kolmogorov machines and that therefore the concept of a 
Kolmogorov machine can be regarded as an adequate formal 
characterization of the concept of an algorithm (see also~\cite{US81a}).

In~\cite{BDG09a}, it is argued that the intuitive notion of algorithmic 
equivalence of programs cannot be captured by an equivalence relation.
This is also argued in the philosophical discussion of the view that
algorithms are mathematical objects presented in~\cite{Dea07a}.
The given arguments are no reason for us to doubt the usefulness of 
studying equivalence relations that capture algorithmic equivalence to a 
certain degree.  
After the appearance of the first version of the current paper, we have 
looked for such equivalence relations.
The results of that search are presented in~\cite{BM14a}.

\bibliographystyle{splncs03}
\bibliography{IS}

\end{document}